\documentclass[fleqn]{article}
\usepackage{longtable}
\usepackage{graphicx}

\begin {document}
\begin{flushleft}
{\LARGE
{\bf Electron Impact Excitation of S~III: an Assessment}
}\\

\vspace{0.5 cm}
[Atoms ({\bf 7}) 2019  --  in press]

\vspace{1.5 cm}

{\bf {Kanti  M  ~Aggarwal}}\\ 

\vspace*{1.0cm}

Astrophysics Research Centre, School of Mathematics and Physics, Queen's University Belfast, \\Belfast BT7 1NN, Northern Ireland, UK\\ 
\vspace*{0.5 cm} 

e-mail: K.Aggarwal@qub.ac.uk \\

\vspace*{0.20cm}

Received: 11 July 2019; Accepted: 15 August 2019

\vspace*{1.0 cm}

\vspace*{1.0 cm}
\vspace*{1.0 cm}

\hrule

\vspace{0.5 cm}

\end{flushleft}

\clearpage


\begin{abstract}

In a recent paper, Tayal et al. [{\em Astrophys. J. Suppl.}    {\bf 2019}, {\emph 242},   9] have reported results for  energy levels, radiative rates (A-values)  and effective collision strengths ($\Upsilon$) for transitions among the 198 levels of Si-like S~III. For the calculations they have  adopted the multi-configuration Hartree-Fock (MCHF) code for the energy levels and A-values, and B-spline $R$-matrix (BSR) code for $\Upsilon$. Their reported results appear to be accurate for energy levels and A-values, but not for $\Upsilon$. Through our independent calculations by adopting the flexible atomic code (FAC), we demonstrate that their reported results for $\Upsilon$ are underestimated, by up to a factor of two, and at all temperatures, particularly for the allowed transitions, but some forbidden ones too.   Additionally, for transitions involving the higher levels the behaviour of their $\Upsilon$ results is not correct.

\end{abstract}

\clearpage

\section{Introduction}
Si-like S~III is an important ion for the studies of a variety of astrophysical plasmas, such as H~II regions, planetary atmospheres and stellar objects -- see Tayal et al. \cite{sst} and references therein. Many of its observed lines varying from infrared to extreme ultraviolet regions have been useful for electron density and temperature diagnostics for which atomic data, including energy levels, radiative rates (A-values), collision strengths  ($\Omega$),  and effective collision strengths ($\Upsilon$),  are required. Realising its importance there have been several studies for its atomic data, mainly by Galav\'{i}s et al. \cite{gmz}, Tayal and Gupta \cite{tg}, Hudson et al. \cite{hrs}, and Grieve et al. \cite{grh}. However, all of these works (and a few more as listed in \cite{sst} and also compiled by the NIST (National Institute of Standards and Technology) team, and available at their website:  {\tt {http://www.nist.gov/pml/data/asd.cfm}}) have some deficiencies and some limitations, as discussed in a more recent paper by Tayal et al. \cite{sst}, who have performed yet another calculation with a larger number of 198 levels of the 3s$^2$3p$^2$, 3s3p$^3$, 3p$^4$, 3s$^2$3p3d, 3s$^2$4$\ell$, (3s$^2$3p) 5s/5p/5d, 3s$^2$3p6s, 3s3p$^2$3d, 3s3p$^2$4$\ell$, and 3s3p$^2$5s configurations, i.e. 18 in total. 

For the generation of wavefunctions, i.e. for the calculations of energy levels and A-values, Tayal et al. \cite{sst} have adopted non-orthogonal orbitals in the multi-configuration Hartree-Fock (MCHF) code, which allows independent correlations  for each level with many orbitals and their configurations. As a result of this their calculated energies and A-values are apparently  fairly accurate when compared with the existing theoretical and experimental data -- see their tables 1 and 2. With these wavefunctions they have carried out further calculations of collisional data, i.e. $\Omega$ and $\Upsilon$, which are required in the analysis of plasmas. Apart from improving the accuracy of energies and A-values, the additional advantage of these non-orthogonal orbitals is the avoidance of pseudo-resonances, which are often a problem in the standard $R$-matrix calculations with orthogonal orbitals. These (unphysical) pseudo-resonances in the variation of $\Omega$, which arise at energies above thresholds due to the orthogonality conditions imposed on the orbitals, included in the generation of wavefunctions but not in the collisional calculations, need to be smoothed over as shown in figure~1 of Aggarwal and Hibbert \cite{ah} for three  transitions of O~III.

Since the adopted  wavefunctions of  Tayal et al. \cite{sst}  are accurate, subsequent calculations for collisional data are also `expected' to be equally accurate, particularly when: (i)  they have included a large range of partial waves with angular momentum $J \le$ 23.5, to ensure convergence of $\Omega$ for most forbidden transitions, (ii) have included the contribution of higher neglected partial waves through a top-up procedure, to ensure the convergence of $\Omega$ for the allowed ones as well, (iii) have considered a large range of energy, up to 12.6~Ryd, and have even extrapolated their values of $\Omega$ for higher energies to determine the values of $\Upsilon$ up to T$_e$ = 10$^6$~K, as accurately as possible, (iv) have resolved resonances in a very fine energy mesh of 0.0001~Ryd, to take care of almost all resonances without giving undue weightage to their widths, and finally (v) have performed relativistic calculations with the B-spline $R$-matrix (BSR) code. Therefore, there should be no reason to suspect their calculations of $\Upsilon$  for accuracy, which has been estimated to be about 20\% for most transitions. Unfortunately, we have a different opinion as discussed below.

Assessing atomic data for accuracy is a (very) difficult task \cite{atom}, particularly for $\Upsilon$. This is because the corresponding experimental results for most transitions of ions are almost non existent and a large calculation, as performed by Tayal et al. \cite{sst}, can not be easily repeated as it requires enormous computational resources,  time and expertise. However, it also does not (necessarily) mean that nothing can be done and one has to accept the results at the face value. In general, forbidden transitions are appreciably affected by the presence of numerous closed-channel (Feshbach) resonances (see for example, figure~1 of Tayal et al.), by more than an order of magnitude in some instances, and particularly towards the lower range of electron temperatures.  For this reason it is much more difficult to assess their accuracy. However, the allowed transitions, particularly the strong (electric dipole) ones, i.e. with significant magnitude of oscillator strengths (f-values), do not show any appreciable resonances, and as a result of it their values of $\Omega$ and $\Upsilon$ vary smoothly with increasing energy and temperature, respectively. Therefore, it is comparatively much easier to assess the accuracy for such transitions. Furthermore, values of $\Omega$ (and subsequently $\Upsilon$) for allowed transitions directly depend on their energy ($\Delta$E) and f-values, because $\Omega_{ij} \sim 4~\omega_{i}(f/\Delta$E$_{ij}$)~ln (E), where $\omega$ is the statistical weight. Therefore, we focus (mainly) on such transitions to make an accuracy assessment of the data reported by Tayal et al.

The easiest way to calculate $\Omega$ is by using  the {\em  Flexible Atomic Code}  (FAC) of Gu \cite{fac}, which is available on the website {\tt {https://www-amdis.iaea.org/FAC/}}. This code is based on the {\em distorted-wave} (DW) method, is fully relativistic, is highly efficient to run, and more importantly, produces results (for background $\Omega_B$) which are comparable with other methods, such as the $R$-matrix, as has been demonstrated in several of our earlier papers for a wide range of ions -- see for example, figure 2 of Aggarwal and Keenan \cite{mgv} for Mg~V. Therefore, to perform our calculations for $\Omega$ and $\Upsilon$ we have adopted this code, and have considered the same 18 configurations as included by Tayal et al. \cite{sst}, and exclusively listed above. However, these configurations generate 346 levels in total, out of which 148 (mainly the higher lying ones) have been omitted by Tayal et al. for computational reason. Additionally, this code does not (automatically) calculate resonances, as the $R$-matrix does, but it should not affect the comparisons as our main interest is in (strong) allowed transitions. 

\begin{table}
\caption{Comparison of  energies ($\Delta$E, Ryd) and oscillator strengths (f-values, dimensionless) for some transitions of S~III. $a{\pm}b \equiv$ $a\times$10$^{{\pm}b}$.} 
\begin{tabular}{rrllllll}  \hline
  I   &    J & Lower Level                & Upper Level                  & $\Delta$E (MCHF) & $\Delta$E (FAC) & f (MCHF) & f (FAC) \\
 \hline
  1  &  25 &  3p$^2$~$^3$P$_0$   &  3p3d~$^3$D$^o_1$                        &  1.351  &  1.488  &  1.392        &  1.487         \\
  2  &  26 &  3p$^2$~$^3$P$_1$   &  3p3d~$^3$D$^o_2$                        &  1.323  &  1.488  &  1.041        &  1.190         \\
  3  &  27 &  3p$^2$~$^3$P$_2$   &  3p3d~$^3$D$^o_3$                        &  1.346  &  1.483  &  1.372        &  1.430         \\
  4  &  29 &  3p$^2$~$^1$D$_2$   &  3s3p$^3$($^2$D)~$^1$D$^o_2$   &  1.293  &  1.447  &  0.995         &  1.060         \\
  4  &  30 &  3p$^2$~$^1$D$_2$   &  3p3d~$^1$F$^o_3$                        &  1.345  &  1.458  &  1.334         &  1.376         \\
  5  &  31 &  3p$^2$~$^1$S$_0$   &  3p3d~$^1$P$^o_1$                        &  1.272  &  1.395  &  2.643         &  2.856         \\
  3  &  10 &  3p$^2$~$^3$P$_2$   &  3s3p$^3$($^2$P)~$^3$P$^o_2$     &  0.893  &  0.952  &  2.171$-$5  &  2.692$-$2  \\ 
  3  &  11 &  3p$^2$~$^3$P$_2$   &  3s3p$^3$($^2$P)~$^3$P$^o_1$     &  0.894  &  0.952  &  9.733$-$6  &  8.665$-$3  \\
  4  &  13 &  3p$^2$~$^1$D$_2$   &  3p3d~$^1$D$_2^o$                       &  0.834  &  0.908  &  2.378$-$2  &  8.828$-$3  \\
13  &  32 &  3p3d~$^1$D$^o_2$   &  3p4p~$^1$P$_1$                           &  0.588  &  0.560  &  6.895$-$2  &  6.711$-$2  \\
20  &  33 &  3p3d~~$^3$P$^o_1$ &  3p4p~$^3$D$_1$                           &  0.147  &  0.165  &  5.735$-$2  &  2.633$-$2  \\
22  &  33 &  3p4s~~$^3$P$^o_0$ &  3p4p~$^3$D$_1$                           &  0.206  &  0.227  &  2.388$-$1  &  4.463$-$1  \\
\hline	
\end{tabular}
\begin{flushleft}
{\small
MCHF: earlier calculations of Tayal et al. \cite{sst} with the MCHF code  \\
FAC: present calculations with the FAC code   \\
}
\end{flushleft}
\end{table}

\section {Collision strengths and effective collision strengths}

A closer look at table~2 of Tayal et al. \cite{sst} shows that there are six transitions with significant f-values, namely 1--25 (3p$^2$~$^3$P$_0$--3p3d~$^3$D$^o_1$), 2--26 (3p$^2$~$^3$P$_1$--3p3d~$^3$D$^o_2$),  3--27 (3p$^2$~$^3$P$_2$--3p3d~$^3$D$^o_3$), 4--29 (3p$^2$~$^1$D$_2$--3s3p$^3$($^2$D)~$^1$D$^o_2$), 4--30 (3p$^2$~$^1$D$_2$--3p3d~$^1$F$^o_3$), and 5--31 (3p$^2$~$^1$S$_0$--3p3d~$^1$P$^o_1$) -- the {\em indices} used are those of Tayal et al. given in their table~1. In Table~1 we list their transition energies and f-values along with our results with FAC for a ready comparison. Tayal et al. have not reported results for $\Omega$ but our data for these six transitions are shown in Figures~1 and ~2. As expected, values of $\Omega$ vary smoothly and increase with increasing energy. The corresponding results for $\Upsilon$ are shown in Figures~3 and 4 at T$_e$ up to 10$^6$~K. Similar results of Tayal et al. are also included in these figures for comparisons. Since both transition energies and f-values are comparable between the two independent calculations (with MCHF and FAC), the subsequent results for $\Upsilon$ are also expected to be comparable, as already stated. Unfortunately, for all these six transitions (and many more) the $\Upsilon$ values of Tayal et al. are highly {\em underestimated}, by up to a factor of two, and almost at all temperatures, although for a few the differences decrease towards the higher end of the temperature range.  In the absence of the $\Omega$ data of Tayal et al. no direct comparisons/conclusions can be made but we can definitely speculate on the source of discrepancies.

\begin{figure*}
\setcounter{figure}{0}
\includegraphics[angle=-0,width=0.9\textwidth]{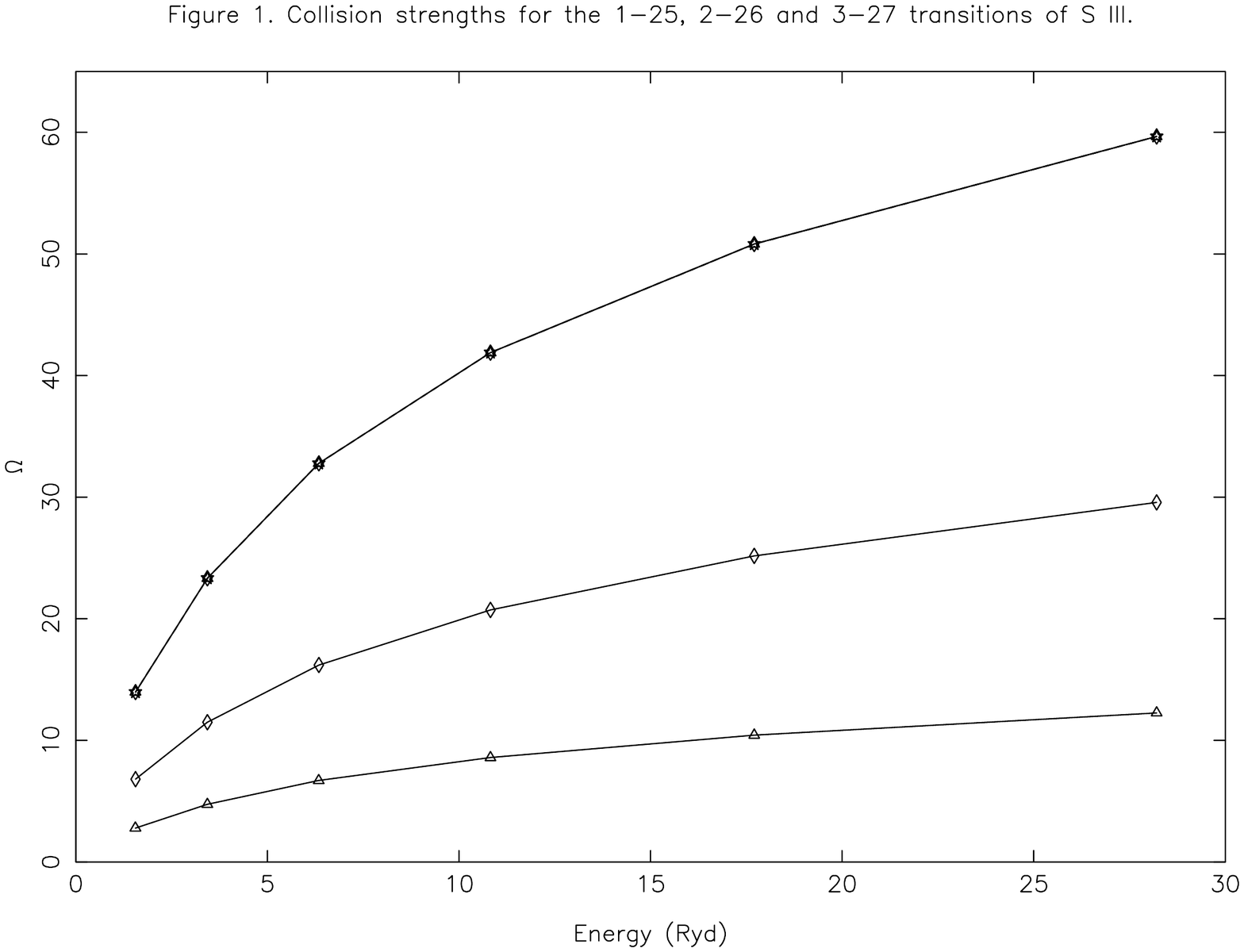}
 \vspace{-1.5cm}
 \caption{Our calculated values of $\Omega$ with FAC for  the 1--25 (triangles: 3p$^2$~$^3$P$_0$--3p3d~$^3$D$^o_1$), 2--26 (diamonds: 3p$^2$~$^3$P$_1$--3p3d~$^3$D$^o_2$), and 3--27 (stars: 3p$^2$~$^3$P$_2$--3p3d~$^3$D$^o_3$) transitions of S~III. }
 \end{figure*}
 
 \begin{figure*}
\includegraphics[angle=-90,width=0.9\textwidth]{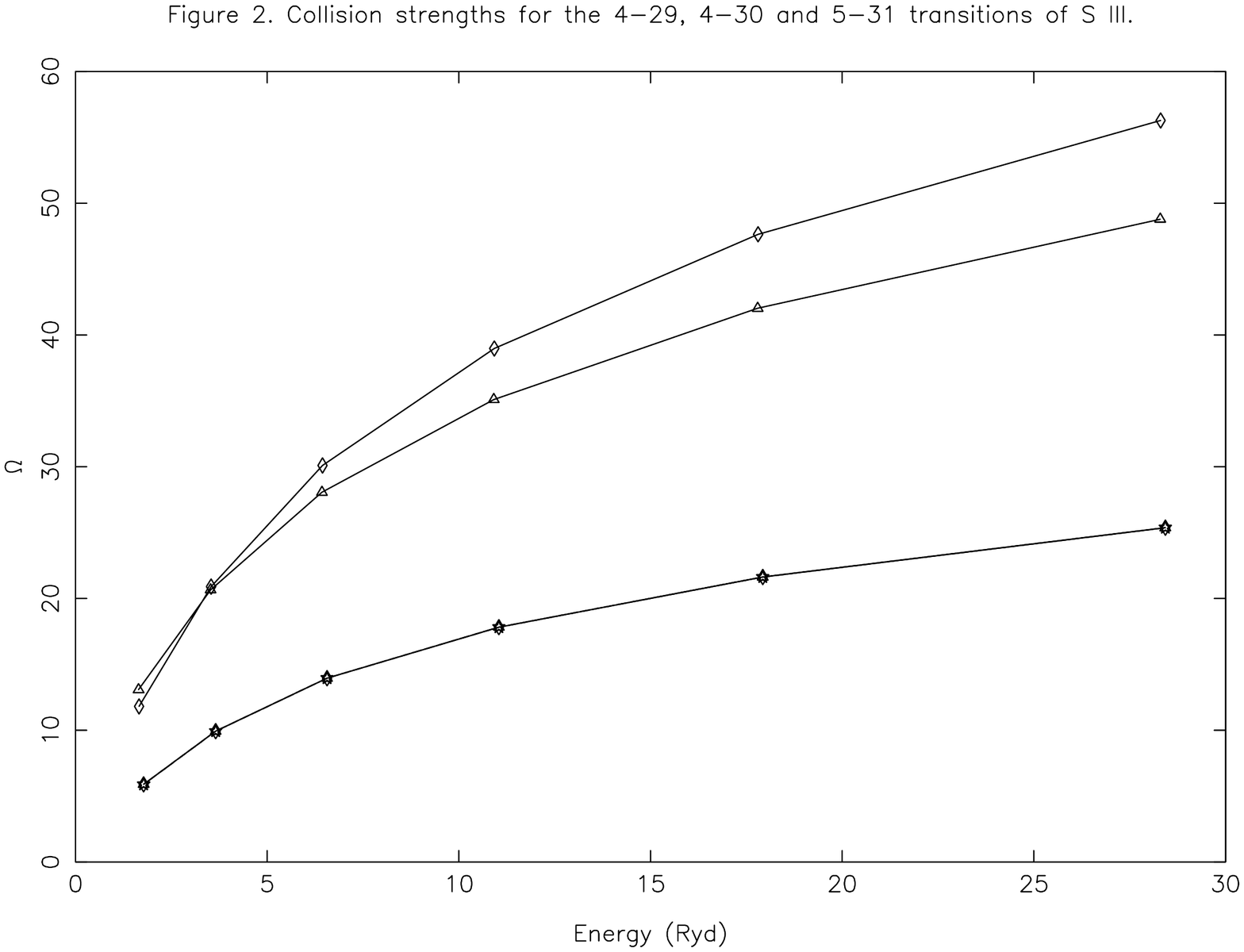}
 \vspace{-1.5cm}
 \caption{Our calculated values of $\Omega$ with FAC for  the 4--29 (triangles: 3p$^2$~$^1$D$_2$--3s3p$^3$($^2$D)~$^1$D$^o_2$), 4--30 (diamonds: 3p$^2$~$^1$D$_2$--3p3d~$^1$F$^o_3$), and 5--31 (stars: 3p$^2$~$^1$S$_0$--3p3d~$^1$P$^o_1$) transitions of S~III.}
 \end{figure*}
 
\begin{figure*}
\includegraphics[angle=0,width=0.9\textwidth]{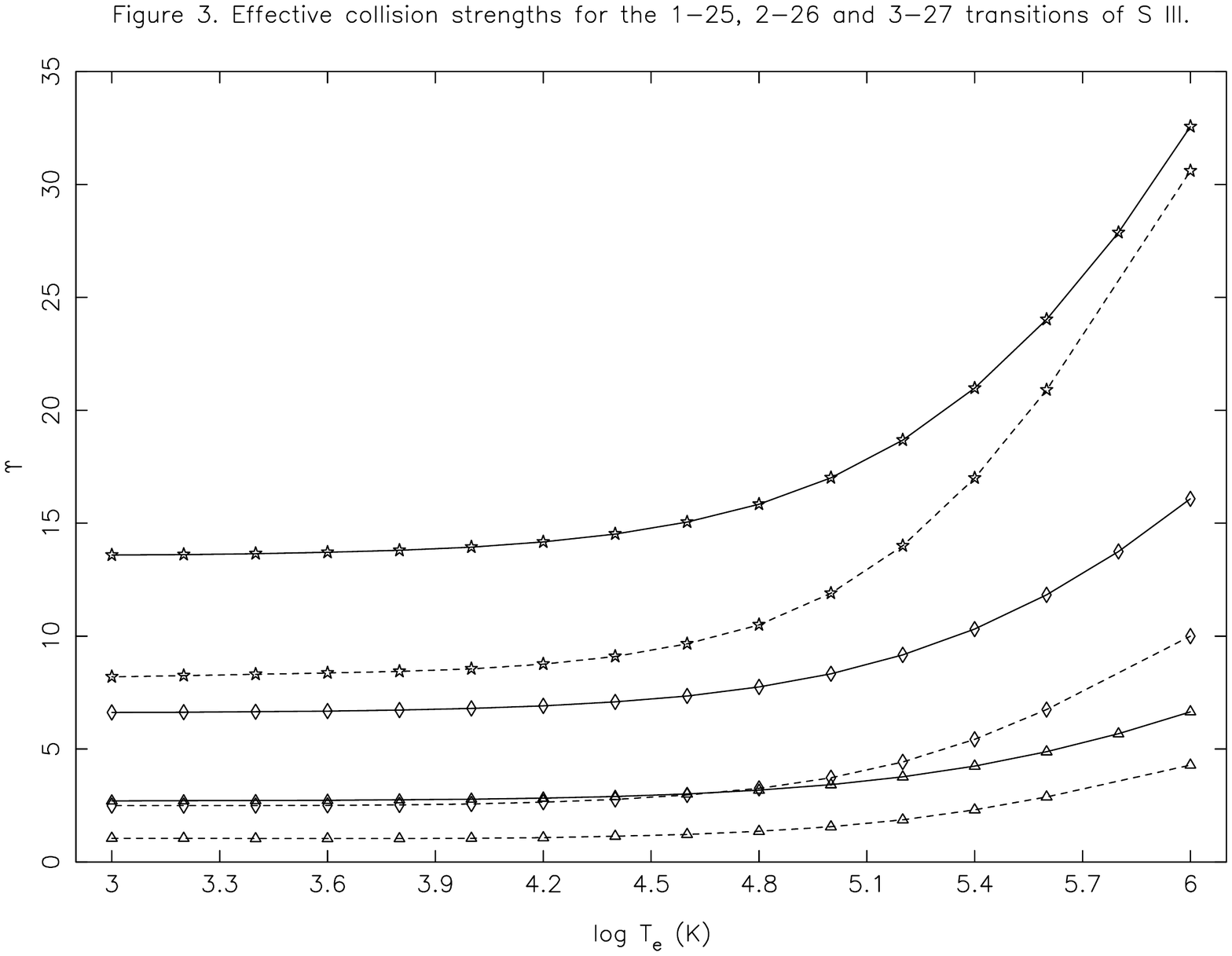}
 \vspace{-1.5cm}
 \caption{Comparison of  FAC and BSR values of $\Upsilon$ for the 1--25 (triangles: 3p$^2$~$^3$P$_0$--3p3d~$^3$D$^o_1$), 2--26 (diamonds: 3p$^2$~$^3$P$_1$--3p3d~$^3$D$^o_2$), and 3--27 (stars: 3p$^2$~$^3$P$_2$--3p3d~$^3$D$^o_3$) transitions of S~III. Continuous curves: present results with FAC, broken curves: earlier results of Tayal et al. \cite{sst} with BSR.}
 \end{figure*}

 \begin{figure*}
\includegraphics[angle=0,width=0.9\textwidth]{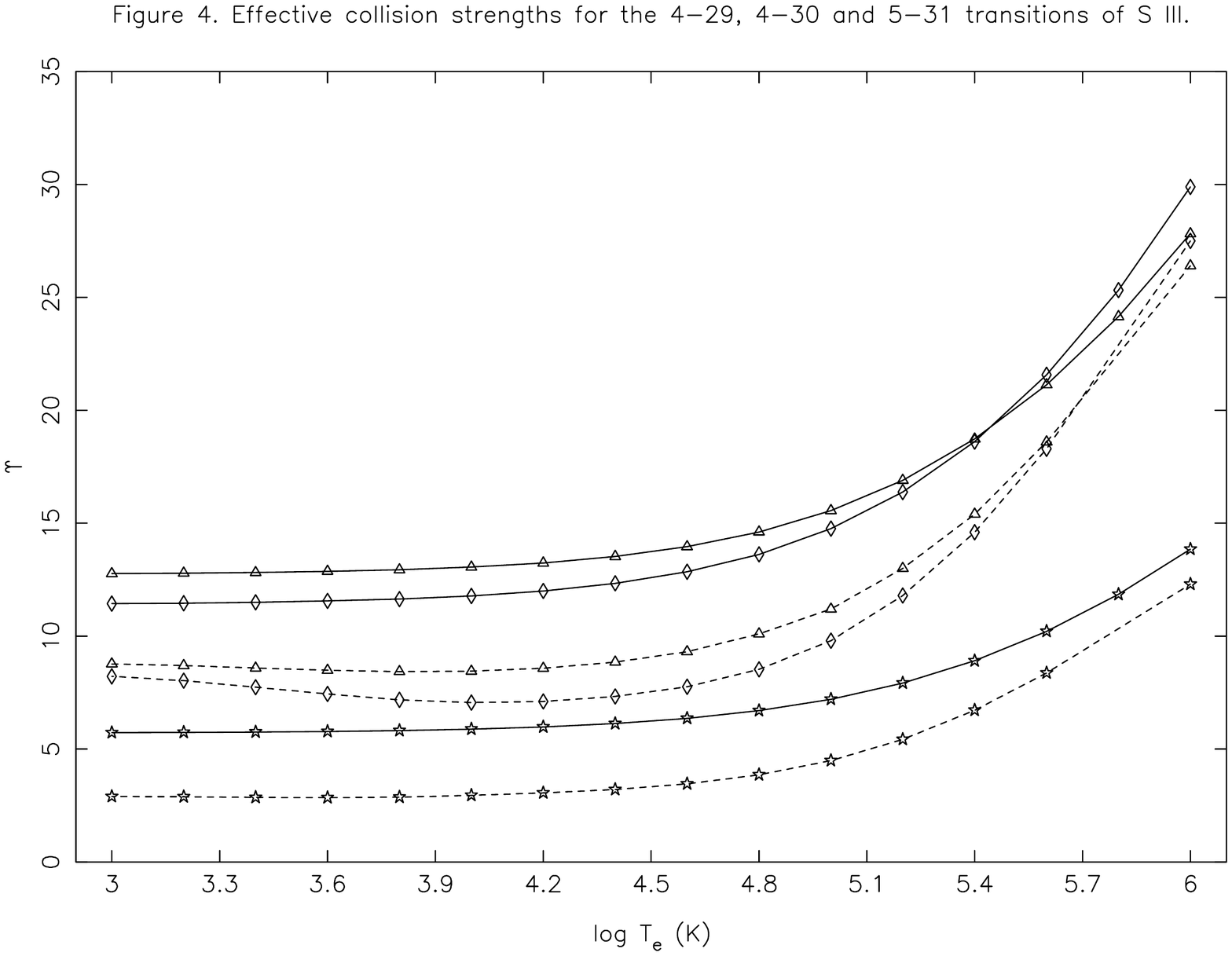}
 \vspace{-1.5cm}
 \caption{Comparison of FAC and BSR  values of $\Upsilon$ for the 4--29 (triangles: 3p$^2$~$^1$D$_2$--3s3p$^3$($^2$D)~$^1$D$^o_2$), 4--30 (diamonds: 3p$^2$~$^1$D$_2$--3p3d~$^1$F$^o_3$), and 5--31 (stars: 3p$^2$~$^1$S$_0$--3p3d~$^1$P$^o_1$) transitions of S~III. Continuous curves: present results with FAC, broken curves: earlier results of Tayal et al. \cite{sst} with BSR.}
 \end{figure*} 
 
\begin{figure*}
\includegraphics[angle=-90,width=0.9\textwidth]{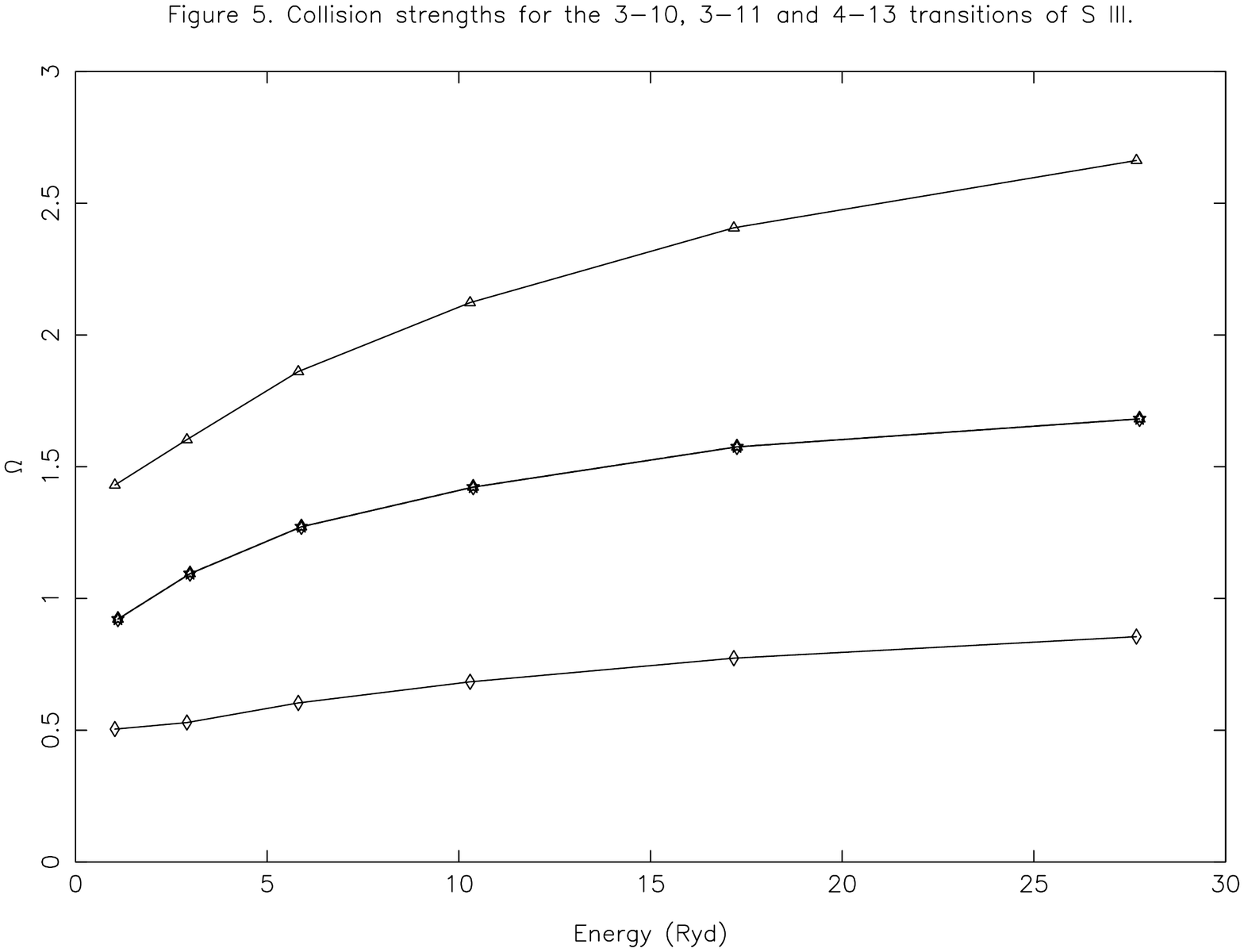}
 \vspace{-1.5cm}
 \caption{Our calculated values of $\Omega$ with FAC for  the 3--10 (triangles: 3p$^2$~$^3$P$_2$--3s3p$^3$($^2$P)~$^3$P$^o_2$), 3--11 (diamonds: 3p$^2$~$^3$P$_2$--3s3p$^3$($^2$P)~$^3$P$^o_1$), and 4--13 (stars: 3p$^2$~$^1$D$_2$--3p3d~$^1$D$_2^o$) transitions of S~III.}
 \end{figure*}
 
\begin{figure*}
\includegraphics[angle=-90,width=0.9\textwidth]{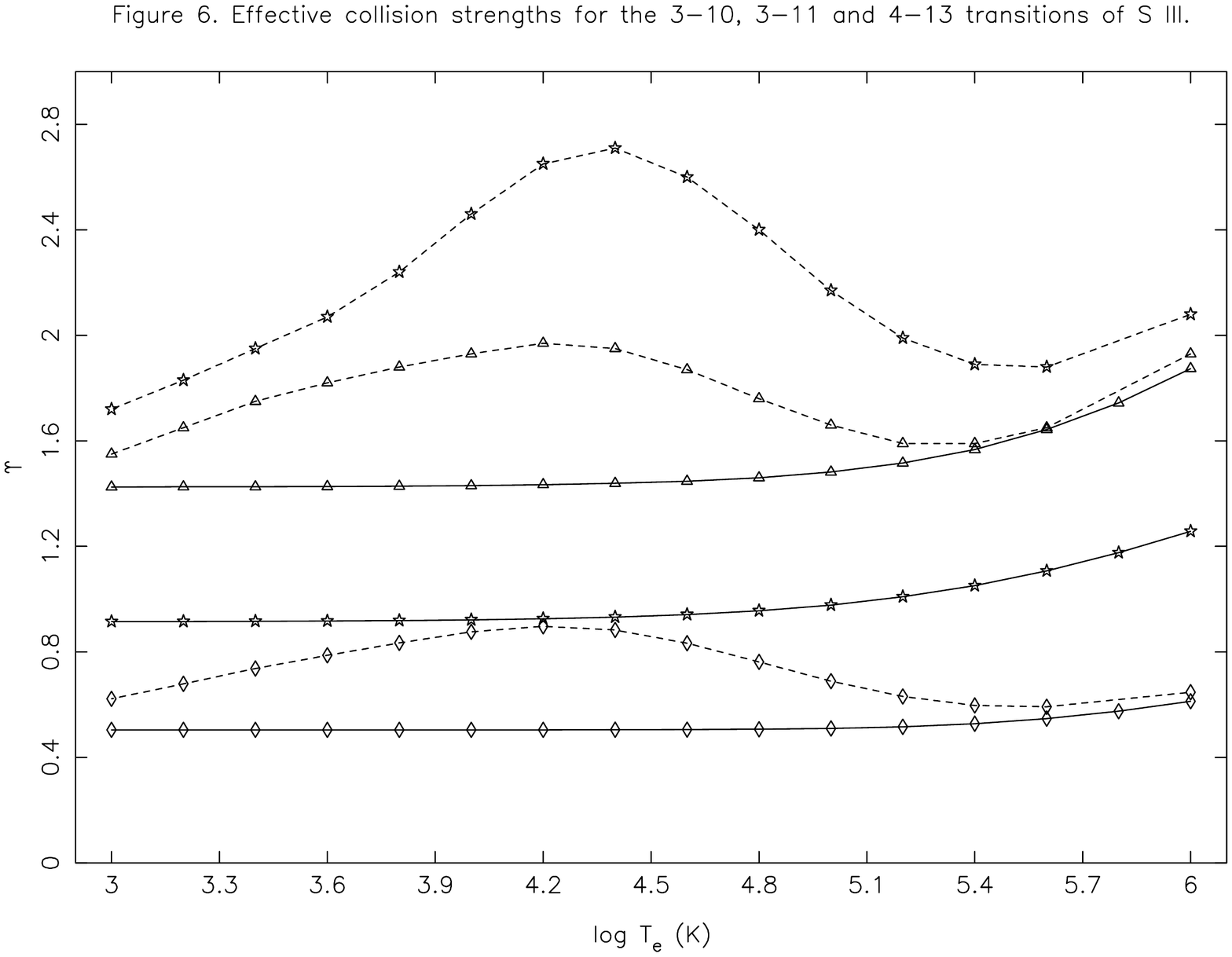}
 \vspace{-1.5cm}
 \caption{Comparison of FAC and BSR  values of $\Upsilon$ for the 3--10 (triangles: 3p$^2$~$^3$P$_2$--3s3p$^3$($^2$P)~$^3$P$^o_2$), 3--11 (diamonds: 3p$^2$~$^3$P$_2$--3s3p$^3$($^2$P)~$^3$P$^o_1$), and 4--13 (stars: 3p$^2$~$^1$D$_2$--3p3d~$^1$D$_2^o$) transitions of S~III. Continuous curves: present results with FAC, broken curves: earlier results of Tayal et al. \cite{sst} with BSR.}
 \end{figure*}

In some $R$-matrix calculations, it is {\em assumed} that $\Omega$ values (for all types of transitions) {\em converge} within a few partial waves, particularly in the thresholds region, and we suspect this is the case with the calculations of Tayal et al. \cite{sst} too. However, there are many transitions (mainly allowed ones but some forbidden also) for which this is not true. As an example, see figure~4 of Aggarwal et al. \cite{fexv} for the 3s3p~$^1$P$^o_1$ -- 3p$^2$~$^1$D$_2$ transition of Fe~XV in which  the whole $\Omega_B$ is lifted upwards with the inclusion of larger ranges of $J$. Similarly, see figure~3 of Aggarwal and Keenan \cite{fexi} for the 3p$^3$3d~$^1$P$^o_1$ -- 3p$^6$~$^1$S$_1$ transition of Fe~XI, which clearly demonstrates the importance of including a larger range of partial waves than (perhaps) done by Tayal et al.  We understand that Tayal et al. have included a top-up to compensate for the higher neglected partial waves, at energies {\em above} thresholds, but this process at an early stage is insufficient, as it underestimates the values of $\Omega$ and has clearly been demonstrated in figure~2 of Aggarwal and Keenan \cite{nixi} for a transition of Ni~XI, in which the effect of top-up is shown at $J \le$  9.5, 19.5, 29.5, and 39.5. Therefore, in our opinion, inclusion of insufficient number of partial waves  is the reason for the underestimated values of $\Upsilon$ reported by Tayal et al. 

Since the above mentioned underestimation of $\Omega$ is mostly at higher energies, one would expect that the values of $\Upsilon$ will only be underestimated towards the higher end of the temperature range, whereas these are at all temperatures, as seen in Figures~3 and 4. This is because the top-up procedure is performed only at energies {\em above} thresholds, and thus the values of $\Omega$ remain {\em uncorrected}\,  in the thresholds region, if sufficient number of partial waves are not included in a calculation. Therefore, all energies and temperatures are affected for $\Omega$ and $\Upsilon$ values. 

In Figures~5 and 6, we consider three other transitions, namely 3--10 (3p$^2$~$^3$P$_2$--3s3p$^3$($^2$P)~$^3$P$^o_2$), 3--11 (3p$^2$~$^3$P$_2$--3s3p$^3$($^2$P)~$^3$P$^o_1$) and 4--13 (3p$^2$~$^1$D$_2$--3p3d~$^1$D$_2^o$), which also have comparable $\Delta$E, as listed in Table~1, but the f-values differ considerably between the MCHF and FAC calculations, as the transitions are {\em weak}. For such transitions, resonances often contribute towards the determination of $\Upsilon$ and this can clearly be noted by the humps seen in Figure~6 in the results of Tayal et al. \cite{sst}. It may also be noted that the BSR values of $\Upsilon$ for the 4--13 transition are larger than our results with FAC because the corresponding f-value is also larger, whereas for the other two transitions the $\Upsilon$ values are similar, particularly towards the higher end of the temperature range,   in spite of the large differences in the f-values. 

In Figure~7 we show our results of $\Omega$ for three {\em forbidden} transitions, namely 4--5 (3p$^2$~$^1$D$_2$--3p$^2$~$^1$S$_0$), 20--25 (3p3d~$^3$P$^o_1$--3p3d~$^3$D$^o_1$) and 20--27 (3p3d~$^3$P$^o_1$--3p3d~$^3$D$^o_3$), which have comparable magnitudes as the ones shown in Figure~5. Particularly for the 4--5 (type of) transition the convergence of $\Omega$ with respect to $J$\,  is slow -- see for example, figure~2 of Aggarwal and Keenan \cite{fexiii} for a similar transition of Fe~XIII. However, for this transition the BSR values of $\Upsilon$ by Tayal et al. \cite{sst} are considerably {\em lower} than our results, and at all temperatures, as seen in Figure~8. This is in spite of the inclusion of resonances in their calculations, and indicates, yet again, that they have included insufficient number of partial waves in their work. Also note that 20--27 is a clear forbidden transition for which both $\Omega$ and $\Upsilon$ should decrease with increasing energy and T$_e$, as is the case in our calculations, but not in the BSR results. However, such small anomalies may sometimes occur in a large calculation, but may not affect the modelling results. 

Finally, in Figures~9 and 10 we show similar results of $\Omega$ and $\Upsilon$, respectively for three other transitions, which are allowed, have comparatively smaller f-values, which do not differ as largely as for the transitions of Figures~5 and 6, but have comparable $\Delta$E (within 10\%), as listed in Table~1. Unfortunately, for these transitions also the $\Upsilon$ values of Tayal et al. \cite{sst} are underestimated at (almost) all temperatures, and this is in spite of their f-value being larger for the 20--33 transition by a factor of two! Therefore, it is clear from the comparisons of $\Upsilon$ shown for 18 representative transitions (including both allowed and forbidden types) that the reported results of Tayal et al. are underestimated, particularly for the (strong) allowed transitions.

 \begin{figure*}
\includegraphics[angle=-90,width=0.9\textwidth]{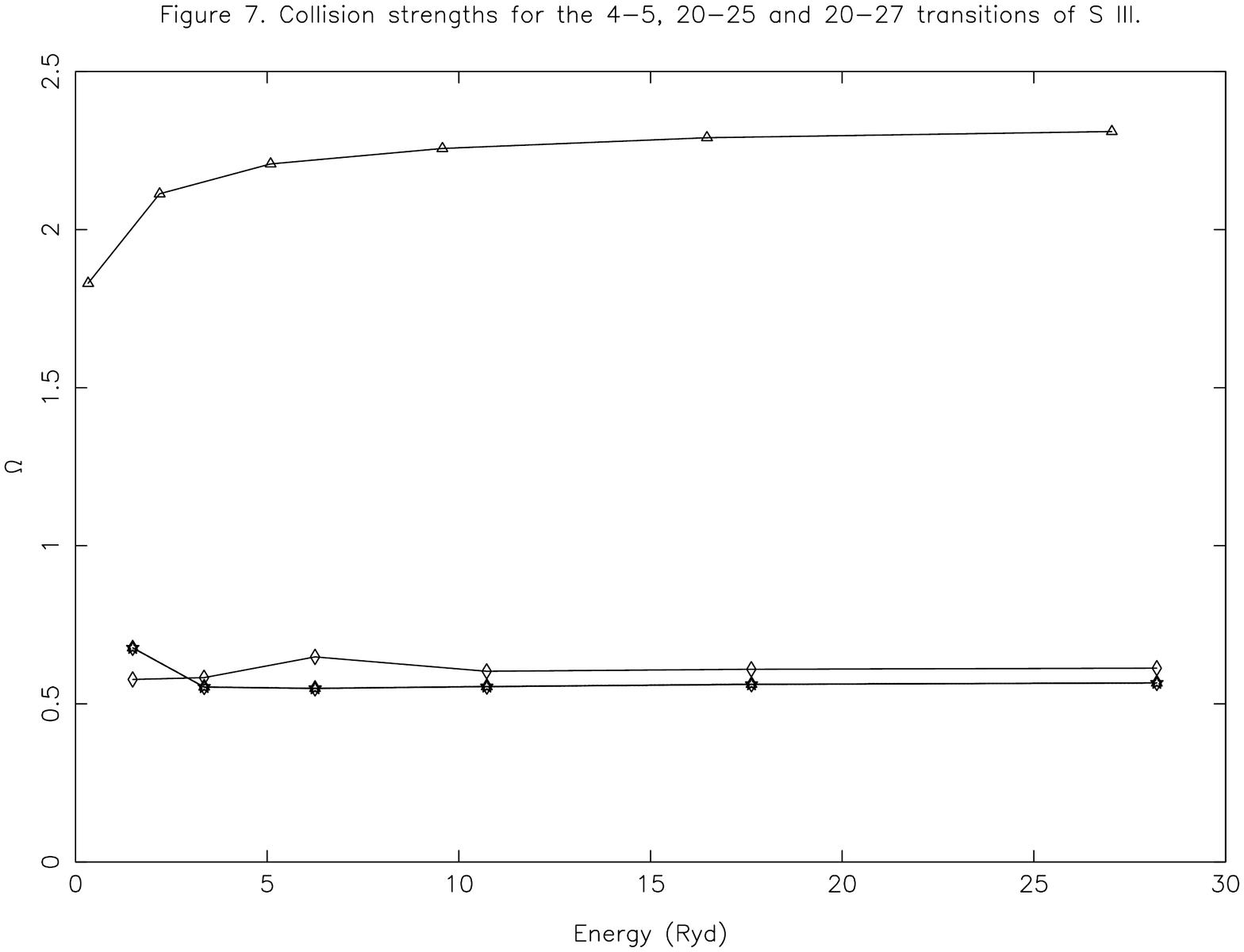}
 \vspace{-1.5cm}
 \caption{Our calculated values of $\Omega$ with FAC for  the 4--5 (triangles: 3p$^2$~$^1$D$_2$--3p$^2$~$^1$S$_0$), 20--25 (diamonds: 3p3d~~$^3$P$^o_1$--3p3d~~$^3$D$^o_1$), and 20--27 (stars: 3p3d~~$^3$P$^o_1$--3p3d~~$^3$D$^o_3$) transitions of S~III.}
 \end{figure*}
 
\begin{figure*}
\includegraphics[angle=0,width=0.9\textwidth]{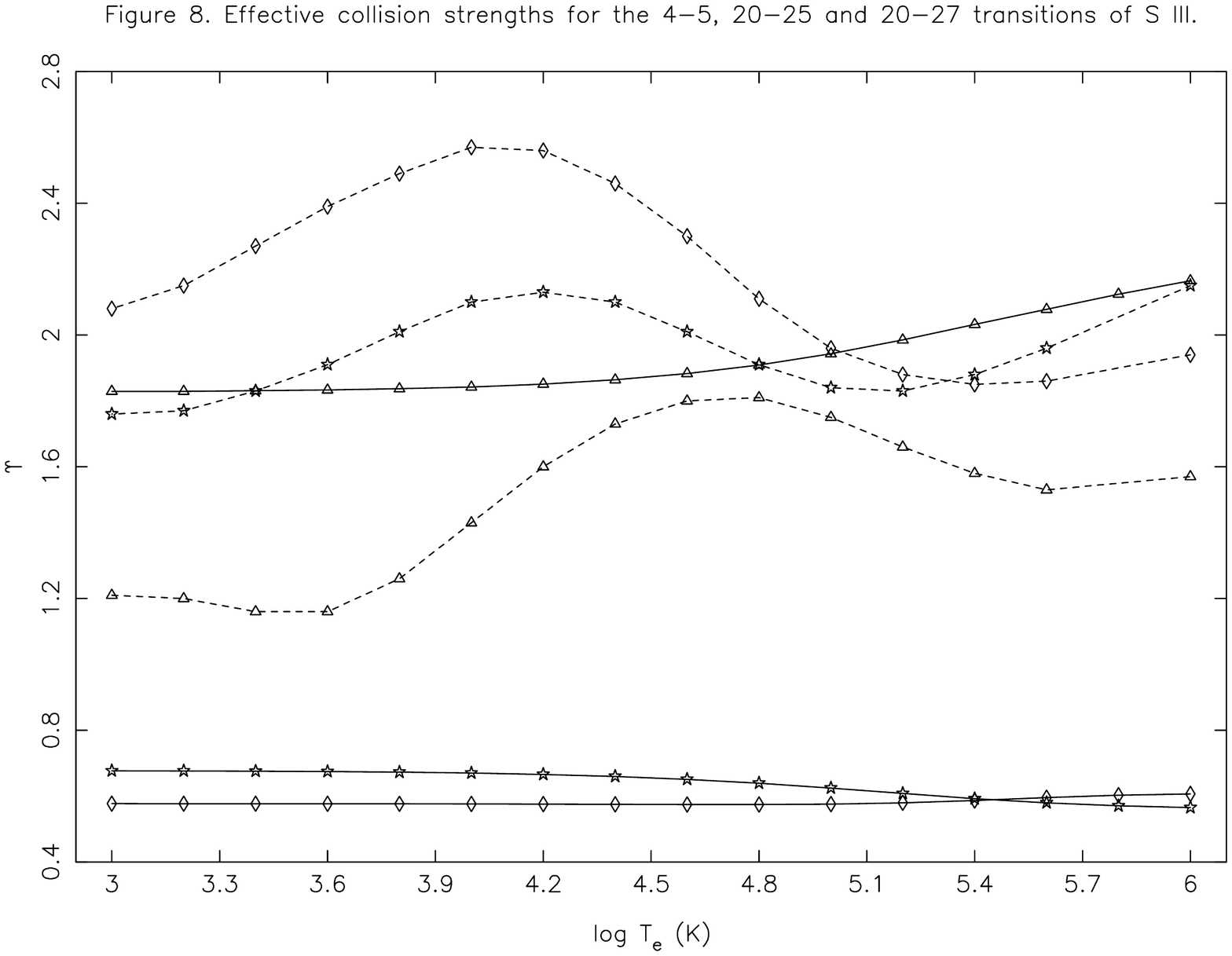}
 \vspace{-1.5cm}
 \caption{Comparison of FAC and BSR  values of $\Upsilon$ for the 4--5 (triangles: 3p$^2$~$^1$D$_2$--3p$^2$~$^1$S$_0$), 20--25 (diamonds: 3p3d~~$^3$P$^o_1$--3p3d~~$^3$D$^o_1$), and 20--27 (stars: 3p3d~~$^3$P$^o_1$--3p3d~~$^3$D$^o_3$) transitions of S~III. Continuous curves: present results with FAC, broken curves: earlier results of Tayal et al. \cite{sst} with BSR.}
 \end{figure*}

\begin{figure*}
\includegraphics[angle=0,width=0.9\textwidth]{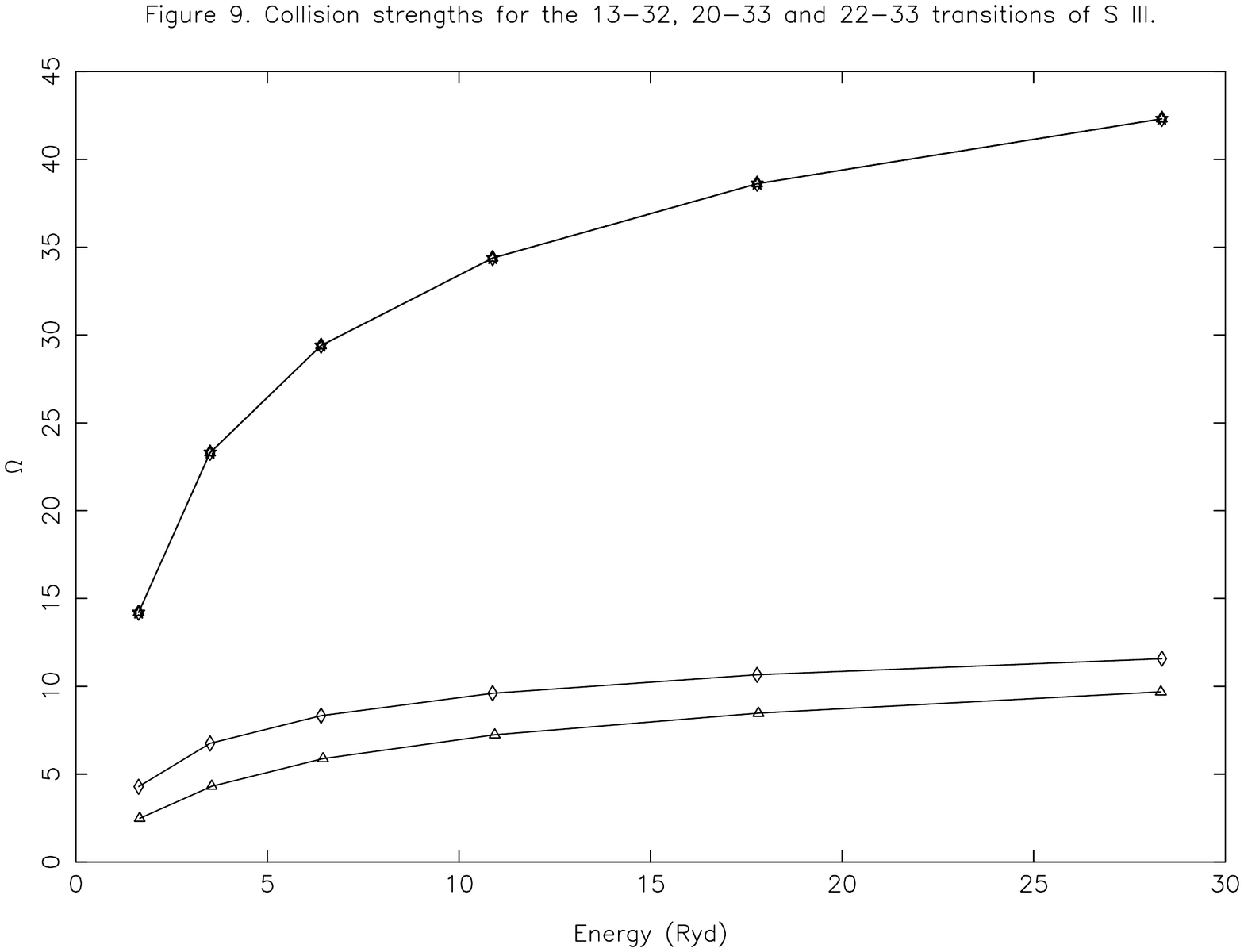}
 \vspace{-1.5cm}
 \caption{Our calculated values of $\Omega$ with FAC for  the 13--32 (triangles: 3p3d~$^1$D$^o_2$--3p4p~$^1$P$_1$), 20--33 (diamonds: 3p3d~~$^3$P$^o_1$--3p4p~$^3$D$_1$), and 22--33 (stars: 3p4s~~$^3$P$^o_0$--3p4p~$^3$D$_1$) transitions of S~III.}
 \end{figure*}
 
\begin{figure*}
\includegraphics[angle=0,width=0.9\textwidth]{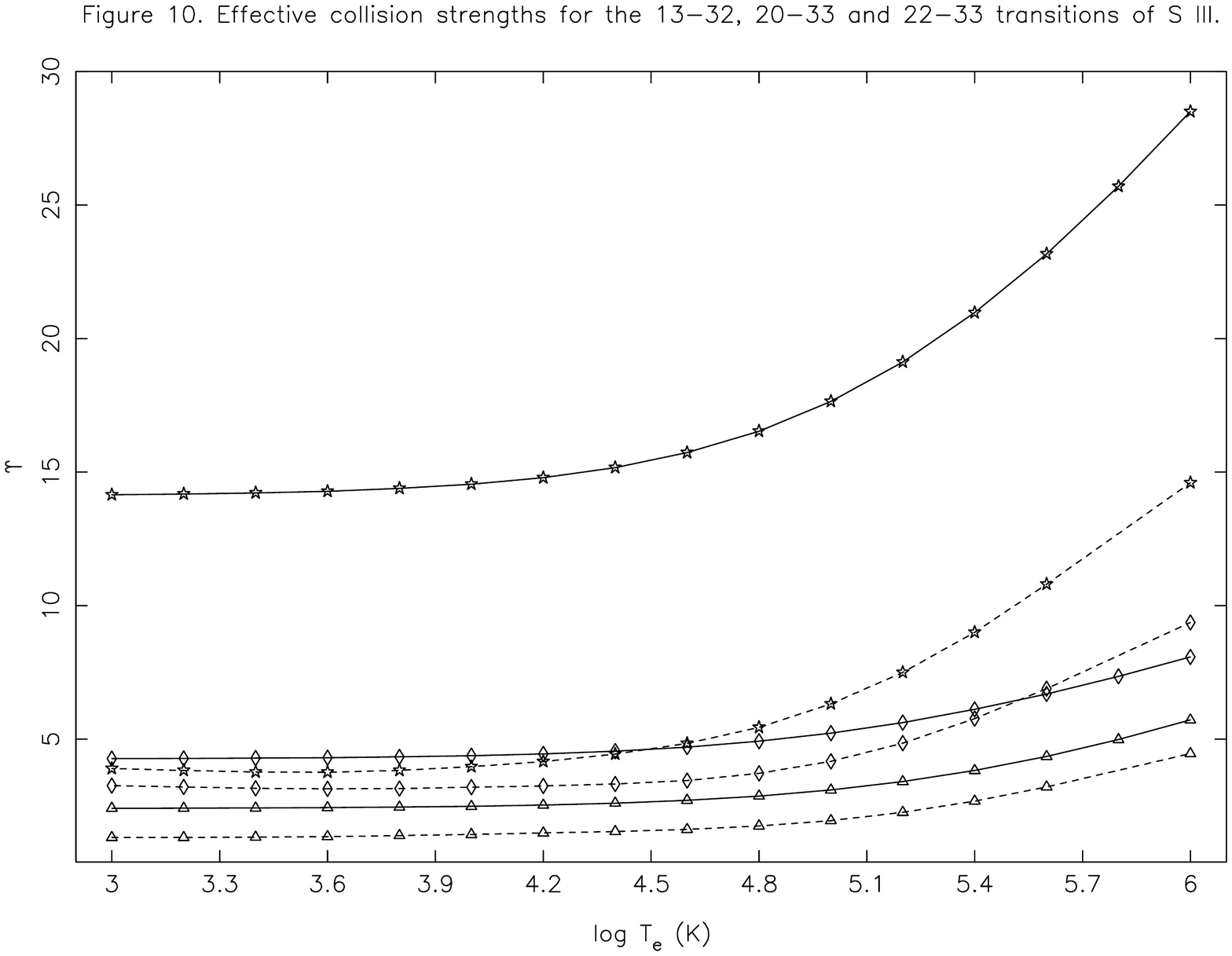}
 \vspace{-1.5cm}
 \caption{Comparison of FAC and BSR  values of $\Upsilon$ for the 13--32 (triangles: 3p3d~$^1$D$^o_2$--3p4p~$^1$P$_1$), 20--33 (diamonds: 3p3d~~$^3$P$^o_1$--3p4p~$^3$D$_1$), and 22--33 (stars: 3p4s~~$^3$P$^o_0$--3p4p~$^3$D$_1$) transitions of S~III. Continuous curves: present results with FAC, broken curves: earlier results of Tayal et al. \cite{sst} with BSR.}
 \end{figure*}

\begin{figure*}
\includegraphics[angle=-90,width=0.9\textwidth]{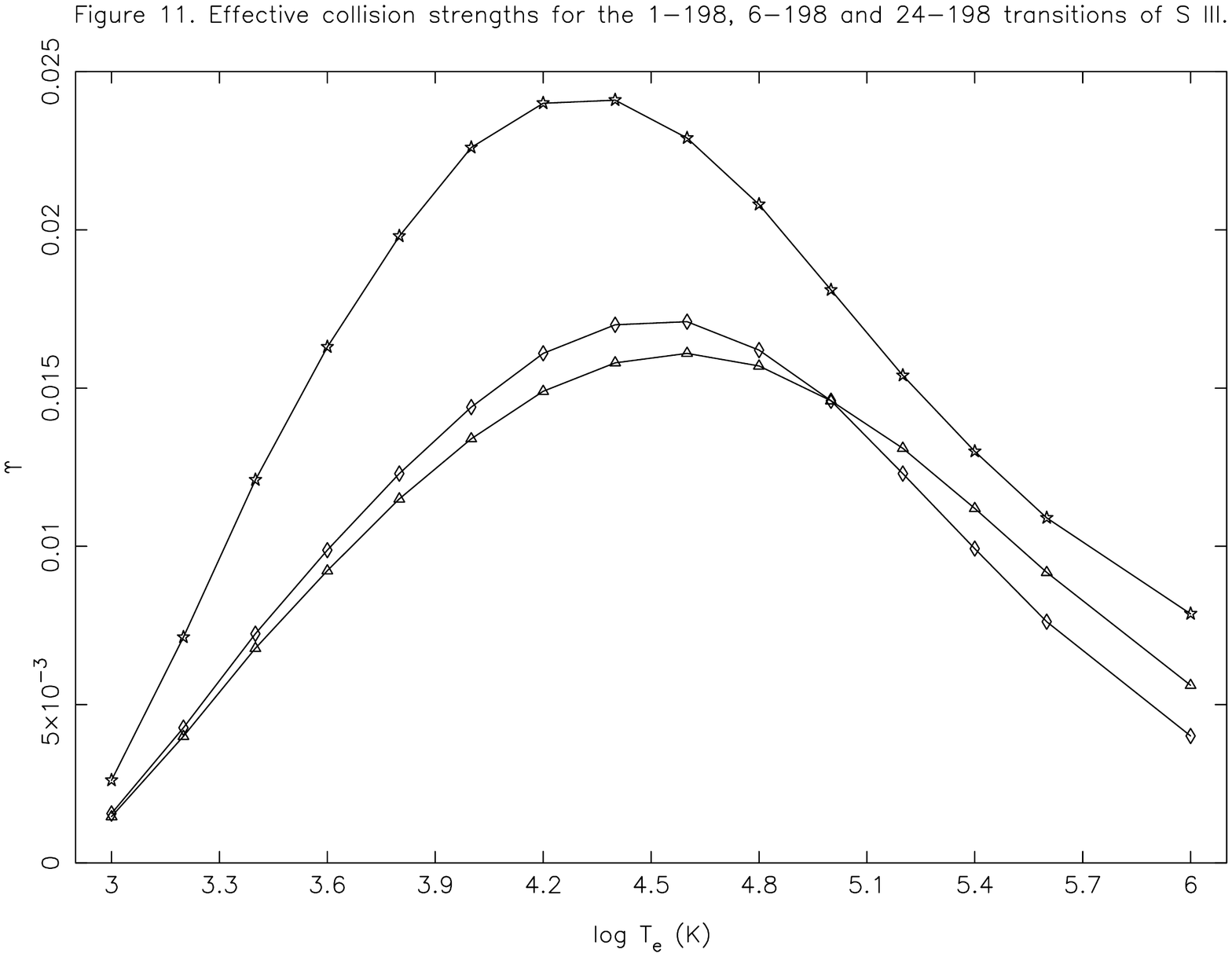}
 \vspace{-1.5cm}
 \caption{The BSR  values of $\Upsilon$ by Tayal et al.  \cite{sst} for the 13--32 (triangles: 3p3d~$^1$D$^o_2$--3p4p~$^1$P$_1$), 20--33 (diamonds: 3p3d~$^3$P$^o_1$--3p4p~$^3$D$_1$), and 22--33 (stars: 3p4s~$^3$P$^o_0$--3p4p~$^3$D$_1$) transitions of S~III.}
 \end{figure*}

\begin{figure*}
\includegraphics[angle=-90,width=0.9\textwidth]{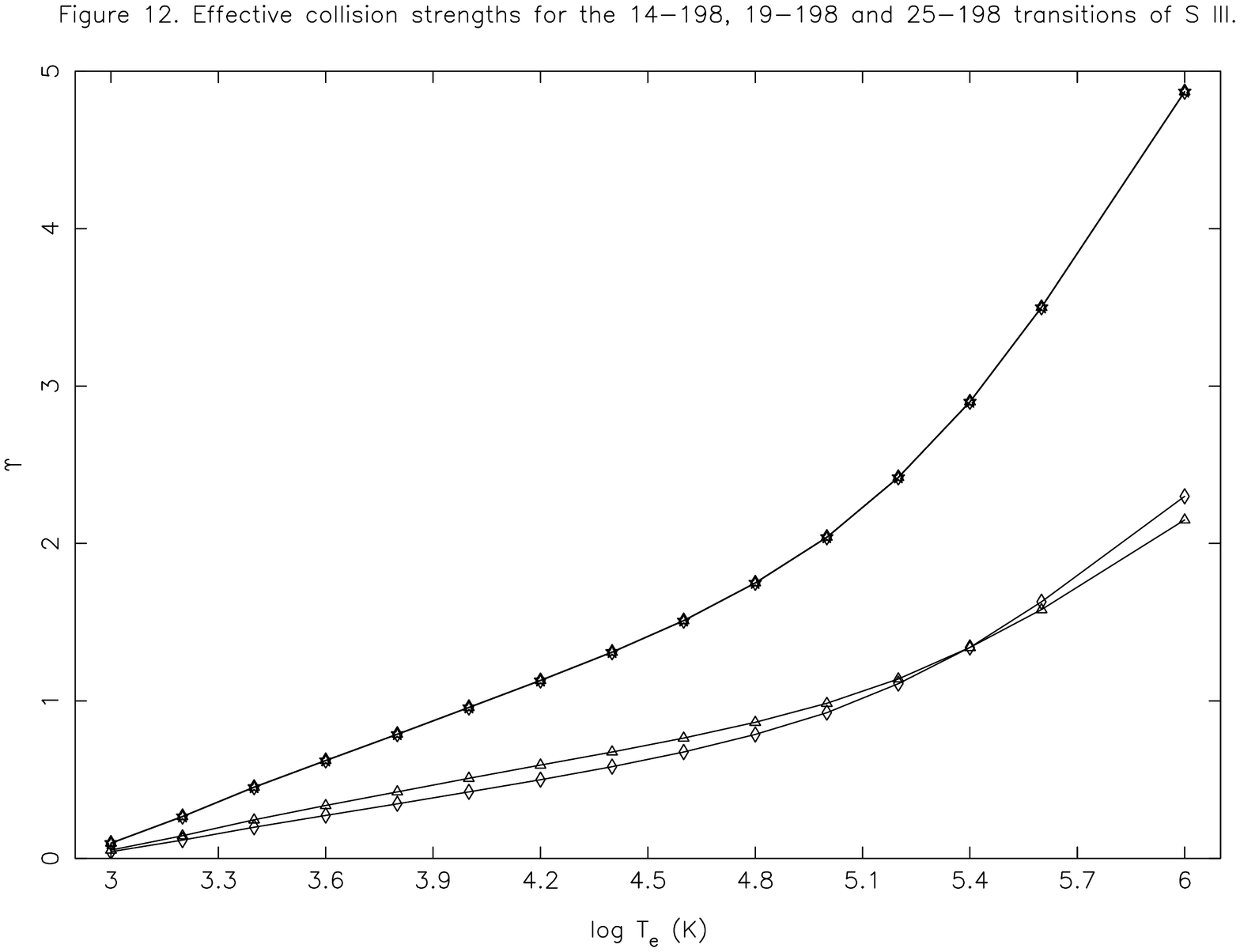}
 \vspace{-1.5cm}
 \caption{The BSR  values of $\Upsilon$  by Tayal et al.  \cite{sst} for the 13--32 (triangles: 3p3d~$^1$D$^o_2$--3p4p~$^1$P$_1$), 20--33 (diamonds: 3p3d~$^3$P$^o_1$--3p4p~$^3$D$_1$), and 22--33 (stars: 3p4s~$^3$P$^o_0$--3p4p~$^3$D$_1$) transitions of S~III.}
 \end{figure*}

Before we conclude, we will like to note that apart from the underestimation of $\Upsilon$ by Tayal et al. \cite{sst} for several transitions of S~III, the behaviour of this parameter is not correct for some transitions involving the higher levels. As an example, in Figure~11 we show the variation of their $\Upsilon$ with T$_e$ for three  transitions which have comparatively small magnitudes, namely 1--198 (3p$^2$~$^3$P$_0$ -- 3s3p$^2$($^2$P)3d~$^3$D$_1$), 6--198 (3s3p$^3$($^4$S)~$^5$S$^o_2$ -- 3s3p$^2$($^2$P)3d~$^3$D$_1$) and 24--198 (3s$^2$3p4s~$^3$P$^o_2$ -- 3s3p$^2$($^2$P)3d~$^3$D$_1$). The first one is parity forbidden whereas the other two are allowed transitions. For any (type of) transition involving the {\em highest} level, there should be no resonances, or equivalently the values of $\Omega$ and hence $\Upsilon$ should vary smoothly, decreasing or increasing, depending on the transition. However, as seen in this figure there are clear ``humps", which indicate the presence of pseudo resonances. In a calculation with non-orthogonal orbitals, as performed by Tayal et al., there should be no  pseudo resonances, and this has also been {\em reaffirmed} by them. This is the same problem as noted earlier (Aggarwal and Keenan \cite{mgv}) on their work on Mg~V -- see also figure~7 of \cite{atom}. However, in a later paper Wang et al. \cite{wang} argued that this behaviour happened due to a ``coarse" energy mesh adopted at energies beyond thresholds, which is 0.2~Ryd in the present case -- also note the contrast between 0.0001~Ryd in the thresholds region and 0.2~Ryd above it. This coarse energy mesh is a cause of the incorrect behaviour seen in the figure. It needs to be finer. We will also like to note that such incorrect behaviours are not confined to the highest level alone, but to others as well --  see for example, 12--195/196/197 transitions. Transitions among the higher lying levels may or may not affect the modelling or diagnostics of plasmas, but the incorrect behaviour of $\Upsilon$ for these certainly undermines confidence in their calculations.

For some other transitions with larger magnitudes, we note a different kind of a problem, shown in Figure~12 for three, namely 14--198 (3s$^2$3p3d~$^3$F$^o_2$ -- 3s3p$^2$($^2$P)3d~$^3$D$_1$), 19--198 (3s$^2$3p3d~$^3$P$^o_0$ -- 3s3p$^2$($^2$P)3d~$^3$D$_1$) and 25--198 (3p3d~$^3$D$^o_1$ -- 3s3p$^2$($^2$P)3d~$^3$D$_1$), all of which are allowed. For these (and several more) transitions the values of $\Upsilon$ are rising too  steeply, i.e. by over a factor of 50 between 10$^3$ and 10$^6$~K, a short span of 6.3~Ryd. Such a steep rise in $\Upsilon$ values is (generally) neither noted nor expected, as may also be confirmed from Figures~3 and 4, beside others. This is yet another indication of their $\Omega$ values not being converged, particularly at the lower end of the energy range. In fact, there are several transitions for which values of $\Upsilon$ increase between the lowest and the highest T$_e$ by up to three orders of magnitude, and examples include 10--141 (3s3p$^3$($^2$P)~$^3$P$^o_2$--3p$^4$~$^3$P$_1$), 11--137 (3s3p$^3$($^2$P)~$^3$P$^o_1$--3s3p$^2$($^4$P)4p~$^3$D$^o_1$), 18--137 (3s3p$^3$($^4$S)~$^3$S$^o_1$--3s3p$^2$($^4$P)4p~$^3$D$^o_1$), 159--198 (3s3p$^2$($^2$D)4p~$^3$D$^o_3$--3s3p$^2$($^2$P)3d~$^3$D$_1$), and 160--198 (3s3p$^2$($^2$D)3d~$^1$D$_2$--3s3p$^2$($^2$P)3d~$^3$D$_1$). Among these only 10--141 is allowed and the rest are forbidden, and particularly for the last two there should be no resonances to contribute, and therefore there is no convincing reason to see such high rise in the values of $\Upsilon$. 

\section{Conclusions}

In this paper, we have made an attempt to assess the recently reported atomic data for S~III by Tayal et al. \cite{sst}. Their calculated energy levels and transition rates appear to be accurate, as claimed. However, their estimated accuracy of 20\% for the collisional parameter ($\Upsilon$), for most transitions, is unrealistic. They have not reported results for $\Omega$ but through our own calculations with FAC we have clearly demonstrated that their reported $\Upsilon$ results are underestimated, by up to a factor of two and (almost) at all temperatures,  for many (strong) allowed transitions as well as (at least a few) forbidden ones too. This is mainly because their included range of partial waves with $J \le$ 23.5 is insufficient for the convergence of $\Omega$, and in the thresholds regions they might  have included even less number of these. Apart from this,  the behaviour of their $\Upsilon$ results is not correct for transitions involving the higher levels, and this is because of the coarse energy mesh adopted at energies above thresholds. 

Adoption of accurate wavefunctions is a desirable criteria for the further calculations of collisional data, which are much more complicated not only to generate but also to assess for accuracy and reliability. Since there is often paucity of corresponding experimental results, accuracy is determined mainly by comparisons with existing results, if available, or by performing other independent calculation by a different method/code, which may not always be possible. Nevertheless,  often large discrepancies are noted among different sets of data for an ion, as discussed, highlighted and explained in one of our papers (Aggarwal \cite{atom}). The assessment of data, particularly for $\Upsilon$ (an important parameter),  becomes even more difficult if the corresponding data for $\Omega$ are not reported. Therefore, we {\em emphasise} yet again that the producers of atomic data should also report results for $\Omega$, at least for some transitions. They may alternatively put these data on an easily accessible website. These steps are very helpful for accuracy assessments and give an opportunity to other workers to make improvements, if possible.

In view of our assessments of available data for $\Upsilon$, we recommend that either fresh calculations be performed for S~III or the authors of \cite{sst} should improve their results by not only enlarging the range of partial waves but by also making their energy mesh finer at energies  above thresholds. Alternatively, they can convincingly demonstrate that their reported results are still accurate and reliable, and our assessment is not satisfactory. Till this is done, our recommendation to the users of data is that they should adopt both calculations, i.e. the one by Tayal et al. \cite{sst} and ours, which can either be easily generated or can be obtained from the author on request, and should then make their own assessment and judgement. In our view the available results of Tayal et al. are (heavily) underestimated for allowed transitions whereas ours for the forbidden ones, because of the omission of resonances. Unfortunately, the two sets of data can also not be (easily) combined because the level orderings are incompatible. 

Our results obtained with FAC have considerable scope for improvements, mainly in two areas. Firstly, our wavefunctions are simple and include only those orbitals and configurations which are also adopted for further calculations of scattering parameters. Inclusion of additional orbitals improves the accuracy of energy levels and A-values, as has been shown by  Tayal et al. \cite{sst}. However, this should not affect the subsequent values of $\Omega$ and $\Upsilon$ by more than 20\%, for most transitions. The other one is the inclusion of resonances, which are omitted in our work. Their contribution is difficult to assess without performing actual calculations, because some forbidden transitions {\em and} at low temperatures may get affected by over an order of magnitude, as may be noted in table~6 of our work \cite{nev} for three transitions of Ne~V.


\end{document}